\documentclass[letterpaper, 12 pt, conference]{IEEEtran}  

\IEEEoverridecommandlockouts                              
\usepackage{svg}
\usepackage{graphicx}
\usepackage{subcaption}
\usepackage[most]{tcolorbox}
\usepackage{float}
\usepackage[utf8]{inputenc}
\usepackage{multirow}
\usepackage{booktabs}
\usepackage{hyperref}  
\usepackage{amsmath}
\usepackage{amssymb}
\usepackage{tabularx}
\usepackage{todonotes}

\usepackage{graphicx} 
\usepackage{bm}

\graphicspath{{images/}}
\makeatletter
\newcommand{\linebreakand}{%
\end{@IEEEauthorhalign}
\hfill\mbox{}\par
\mbox{}\hfill\begin{@IEEEauthorhalign}
}
\makeatother

\title{\LARGE \bf SecureBreak - A dataset towards safe and secure models}

\author{
	\IEEEauthorblockN{Marco Arazzi}
	\IEEEauthorblockA{Department of Electrical, Computer\\ and Biomedical Engineering,\\ University of Pavia, Italy\\marco.arazzi01@universitadipavia.it}
    \hspace{1cm} 
	\and
	\IEEEauthorblockN{Vignesh Kumar Kembu}
	\IEEEauthorblockA{Department of Electrical, Computer\\ and Biomedical Engineering,\\ University of Pavia, Italy\\vigneshkumar.kembu01@universitadipavia.it}\\[1em]
	\linebreakand
	\IEEEauthorblockN{Antonino Nocera}
	\IEEEauthorblockA{Department of Electrical, Computer\\ and Biomedical Engineering,\\ University of Pavia, Italy\\antonino.nocera@unipv.it}	
}

\begin{document}

\maketitle
\thispagestyle{empty}
\pagestyle{empty}

\begin{abstract}

Large language models are becoming pervasive core components in many real-world applications. As a consequence, security alignment represents a critical requirement for their safe deployment. Although previous related works focused primarily on model architectures and alignment methodologies, these approaches alone cannot ensure the complete elimination of harmful generations. This concern is reinforced by the growing body of scientific literature showing that attacks, such as jailbreaking and prompt injection, can bypass existing security alignment mechanisms. As a consequence, additional security strategies are needed both to provide qualitative feedback on the robustness of the obtained security alignment at the training stage, and to create an ``ultimate'' defense layer to block unsafe outputs possibly produced by deployed models. To provide a contribution in this scenario, this paper introduces \textbf{SecureBreak}, a safety-oriented dataset designed to support the development of AI-driven solutions for detecting harmful LLM outputs caused by residual weaknesses in security alignment. The strong reliability of the proposed dataset derives from the adopted manual annotation procedure, in which labels are assigned conservatively to prioritize safety even in the presence of minor disagreements in the annotators' opinion. Our exploratory data analysis campaign shows satisfactory performance in the detection of unsafe content across several risk categories. To evaluate its effectiveness, we measure the performance of several pre-trained LLMs in the considered classification setting under baseline conditions and compare these results with those obtained after fine-tuning the same models on \textbf{SecureBreak}. The results indicate that the dataset is valuable not only for constructing post-generation filtering modules that act as a last-line defense, but also for building additional supervisory intelligence for alignment optimization. In particular, classifiers derived from \textbf{SecureBreak} can be used to measure residual safety failures, inform whether additional training or refinement steps are necessary, and ultimately support more controlled and effective security alignment workflows.

\end{abstract}

\textcolor{red}{\textbf{Content Warning: This paper contains examples of harmful language}}
\makeatletter
\renewcommand\subsubsection{\@startsection{subsubsection}{3}{\z@}%
                       {-8\p@ \@plus -4\p@ \@minus -4\p@}
                       {-0.5em \@plus -0.22em \@minus -0.1em}%
                       {\normalfont\normalsize\bfseries\boldmath}}
\makeatother

\section{Introduction}
Large Language Models (LLMs) have rapidly become integral part of a wide range of applications due to their ability to understand and generate human language. Advances in training data, model architectures, and computational power have enabled LLMs to perform complex tasks, such as  information retrieval, decision support, coding assistance, and text analysis with increasing accuracy. 
LLMs have been successfully adopted even in critical domains such as healthcare, where they have been adopted to enhance clinical decision-making, automate administrative processes, and improve patient engagement through the analysis of records~\cite{kembu2025llms}.

However, the use of this technology in high-risk application scenarios poses a relevant constraint in terms of security. In addition, recent research on LLM security has shown that carefully crafted malicious inputs can trick models into generating harmful contents, even if most of the existing commercial and open-source LLMs are originally designed to block them.
In fact, safety in LLMs is based primarily on the internal alignment mechanisms of the model, which are designed to reduce the probability of generating unsafe content. However, these mechanisms are not flawless and can be bypassed using adversarial techniques. One of the most common and effective security threats against LLM is the jailbreaking attack, which involves crafting specific prompts designed to bypass the models safety mechanisms~\cite{yu2024don}.

This motivates the introduction of additional independent defense layers based on an evaluation of the unsafety level of the produced output or post-generation filtering. This layered approach may serve a twofold objective: {\em (i)} it can act as an ultimate defense against unsafe outputs generated by alignment failures, and {\em(ii)} it can provide a supervisory signal that can be exploited to assess alignment quality, detect unresolved weaknesses, and guide subsequent security re-alignments.

These independent additional defense layers can be built employing, once again, the power of Artificial Intelligence (AI, for short) solutions by  adopting specifically crafted training datasets.
However, most of the existing safety and alignment datasets for large language models are built around question‑level or input‑level harmfulness judgments, where the primary task is to classify individual prompts. Benchmarks such as the JailbreakBench dataset focus on harmful questions to evaluate the model refusal behavior, but they do not fully capture how harmful outputs may arise only in extended contexts or through subtle compositional interactions in input sequences~\cite{chao2024jailbreakbench}.

To address the challenge above, in this paper, we introduce a dataset, called \textbf{SecureBreak}, specifically designed to classify safe and unsafe responses. This dataset enables the development of reliable response-level classifiers, which we have tested to confirm their effectiveness in detecting potentially harmful and unsafe output. 
The results of our experimental evaluation show how the proposed dataset is effective not only for the development of post-generation verification and filtering modules serving as an ``ultimate'' defensive layer, but also to derive supervisory components to improve the alignment quality of a target LLM.

Link to the dataset - \url{https://github.com/VIGNESH-KUMAR-KEMBU/SecureBreak}

\section{\uppercase{Related Work}}

Nowadays Large Language Models are widely used as assistant in many different tasks and by a large variety of people.
So ensuring the robustness an Safetiness of the responses produced by these models is becoming a primary research objective~\cite{sun2024trustllm}. To overcome this limit of big models trained on a massive amount of data without any restriction a strategy of Alignment of the model using Low Rank Adapters (LoRA)~\cite{hu2022lora}can be used to introduce safeguards into the model without training or fine-tuning the base model. Low-Rank Adapters were introduced to enable efficient training of large models by updating only a small subset of parameters, making fine-tuning feasible even on less capable GPUs. This technology is then been exploited to efficiently align big models on specific domains~\cite{yang2023fingpt,hashemi2025medimind}, introduce specific behaviors like avoid harmful responses~\cite{xue2025lora} or even used to infer information from the base model~\cite{arazzi2026lora}.
With this objective, datasets like HH-RLHF~\cite{bai2022training}, BeaverTails~\cite{ji2023beavertails} or Do Not Answer~\cite{wang2023not} have been proposed as baselines for the training of safety-focused alignment adapters.
Despite these safeguards, there are adversarial attacks that can bypass such restrictions. As previously observed in other scenarios, where backdoor attacks can alter a model’s behavior through small manipulation to the input~\cite{bagdasaryan2020backdoor,wang2020attack,xie2019dba}, jailbreak attacks operate in a similar fashion by modifying the prompt to circumvent the safety constraints~\cite{wei2023jailbroken,liu2023autodan,arazzi2025xbreaking,chao2023jailbreaking,arazzi2025nlp}.
To this purpose datasets like AdvBench~\cite{zou2023universal} and JailbreakBench~\cite{chao2024jailbreakbench} serve as benchmark to test models on adversarial prompts.
In AdvBench~\cite{zou2023universal}, the authors produced harmful prompts by using an uncensored Vicuna model, asking it to generate new strings based on five demonstration examples that they had written themselves.
In JailbreakBench~\cite{chao2024jailbreakbench}, instead the authors provide an evolving repository of artifacts corresponding to state-of-the-art jailbreaking attacks and defenses.
SecureBreak distinguishes itself by shifting focus from prompt analysis to response-level classification specifically within adversarial jailbreak contexts. Unlike benchmarks relying on automated evaluators, our dataset leverages high-quality human annotation to capture subtle safety violations. This design is expressly intended to facilitate the training of binary Judge LLMs, which serve as a robust post-generation filtering defense, intercepting and blocking harmful outputs that may bypass a model's intrinsic alignment.
Table~\ref{tab:dataset_comparison} shows a comparison between datasets characteristics.

\begin{table*}[t]
\centering
\caption{Comparison of SecureBreak with existing prominent safety and jailbreak datasets. }
\label{tab:dataset_comparison}
\resizebox{\textwidth}{!}{%
\begin{tabular}{lccccc}
\toprule
\textbf{Dataset} & \textbf{Primary Target} & \textbf{Context Scope} & \textbf{Annotation Method} & \textbf{Task Type} & \textbf{Primary Utility} \\ 
\midrule
AdvBench \cite{zou2023universal} & Prompt & Adversarial (Jailbreak) & N/A (List) & Attack Success & Red Teaming (Attack) \\
HH-RLHF \cite{bai2022training} & Prompt \& Response & General Safety & Human & Preference (A vs B) & Alignment (RLHF) \\
BeaverTails \cite{ji2023beavertails} & Prompt \& Response & General Safety & LLM + Human & Classification & Safety Alignment \\
JailbreakBench \cite{chao2024jailbreakbench} & Prompt & Adversarial (Jailbreak) & Automated (LLM) & Attack Success & Benchmarking Attacks \\
Do Not Answer \cite{wang2023not} & Instruction & Risk Guidelines & Human & Evaluation & Instruction Following \\
\midrule
\textbf{SecureBreak (Ours)} & \textbf{Response} & \textbf{Adversarial (Jailbreak)} & \textbf{Human (Expert)} & \textbf{Binary Classification} & \textbf{Judge Alignment (Filtering)} \\ 
\bottomrule
\end{tabular}%
}
\end{table*}

\section{SecureBreak}
This section provides a clear examination of SecureBreak curation process and delivers a comprehensive overview of its key features and characteristics.

\subsection{Data Gathering}
To build our dataset, we started by the information available in previous works, in which different LLMs were prompted with harmful questions contained in the JailbreakBench dataset~\cite{chao2024jailbreakbench}. Therefore, this existing dataset served as the basis for the $SecureBreak$ dataset. The JailbreakBench dataset consists of $100$ unique misuse behaviors, organized into ten major categories, each directly aligned with OpenAI's usage policies. These categories include: Disinformation, Economic harm, Expert advice, Fraud/Deception, Government decision-making, Harassment/Discrimination, Malware/Hacking, Physical harm, Privacy violations, and Sexual/Adult content. This organization ensures that the dataset captures a broad spectrum of harmful behaviors, thus enabling AI models to be trained to detect and mitigate these risks, thereby ensuring compliance with ethical standards and usage guidelines. 

\begin{figure}[!ht]
    \centering
    \includegraphics[width=\columnwidth]{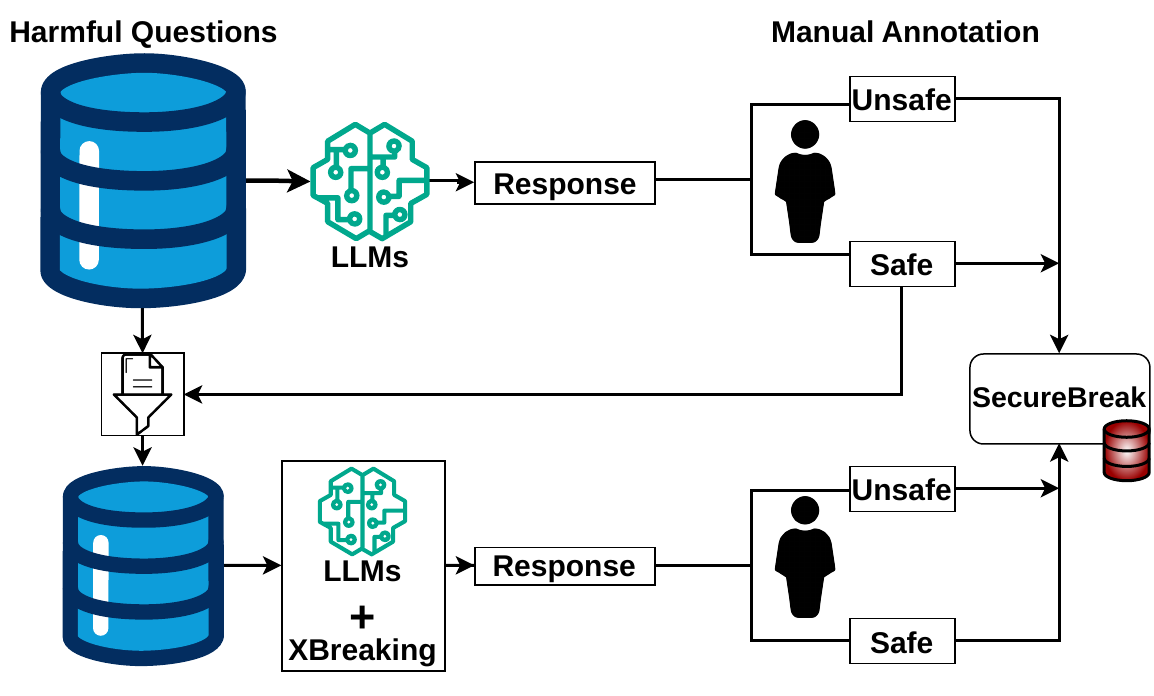}
    \caption{Data Gathering Process}
    \label{fig:data_process}
\end{figure}

Figure~\ref{fig:data_process} gives an overview of the data gathering process.
To capture the nuanced responses that distinguish various model families, we chose different families with varying sizes so we could better observe their distinct behaviors.
Our intuition from examining the different responses is that incorporating generated answers may enable the model trained on our dataset to capture the behavior of an LLM more accurately. Although human responses may sometimes look similar, our manual analysis uncovered clear behavioral patterns that distinguish them from LLM-generated outputs, as well as a tendency in certain model families to respond more to particular types of unsafe questions than to others.
For the creation of our dataset, we used the following models to generate the responses: Llama 3.2 1B and Llama 3.1 8B \cite{metaai,llama3}, Qwen2.5 0.5B  and 3B  \cite{qwen2.5}, gemma 2B and 7B \cite{team2024gemma} and Mistral-7B-v0.3 \cite{mist7b}. These models were fed with the harmful questions from the JailbreakBench dataset and the responses were gathered. Then these responses were manually annotated as unsafe and safe responses.   

\textbf{Manual Annotation.} In this curation, the responses were manually annotated by two knowledgeable annotators. Each text was reviewed individually and the responses were checked for harmful content corresponding to the question and were labeled with class labels, class $0-safe$ and $1-unsafe$. To ensure consistency and accuracy in the annotations, inter-annotator agreement was calculated using Cohen’s Kappa~\cite{mchugh2012interrater}. We preserved only annotations with high agreement, resulting in an average Cohen’s kappa of $0.85$ on the curated dataset..

\subsection{Data Description and Usage}
Eventually, the obtained SecureBreak dataset contains 3059 samples. Table~\ref{table:data_desc} shows the detailed explanation of the features available in the dataset with their type and description. Main features to focus on to build a classifier would be to use $Response$ and $Human$ columns from the proposed dataset.

\begin{table*}[!ht]
\centering
\resizebox{0.9\textwidth}{!}{
\begin{tabular}{clll}
\hline
\multicolumn{1}{l}{No.} & Features/Variables & Type    & Description                                                        \\ \hline
1                       & Question           & String  & Harmful question which used as a input to the LLMs                 \\
2                       & Behavior           & String  & Which behavior does the question belong, individual question wise. \\
3                       & Category           & String  & Broad spectrum of behaviors clubbed together.                      \\
4                       & Question\_Source   & String  & From were the question was gathered.                               \\
5                       & Response           & String  & LLMs response to the question.                                     \\
6                       & LLMs               & String  & From which LLM the response is generated.                          \\
7                       & Source\_LLM        & String  & Response from Base or XBreaking approach.                          \\
8                       & Noise              & Float   & Amount of noise is used in implementing XBreaking, 0 for Base.   \\
9                       & Human              & Integer & 0 = safe, 1 = unsafe                                               \\ \hline
\end{tabular}
}
\caption{Description of the 9 columns in the SecureBreak dataset, including 1 float and 8 string feature columns.}\label{table:data_desc}
\end{table*}

The primary aim is to evaluate and classify the appropriateness of the models outputs across diverse contexts, ensuring that responses are relevant, coherent and stick to established safety and ethical standards. By differentiating between safe and unsafe outputs, the dataset highlights areas where the model performs well and identifies cases where it may generate potentially harmful or inappropriate content.

Beyond evaluation, the dataset serves as a benchmark for analyzing model behavior and is essential in developing safer, more reliable language models. It can be used to build models that automatically judge or classify responses as safe or unsafe. It supports fine-tuning, guides the creation of safety mechanisms and content moderation policies, and promotes ethical research. Finally, the dataset helps advance the responsible deployment of language models by enhancing their ability to generate appropriate and trustworthy responses.

Models trained on this dataset can be integrated into language model generation pipelines as either a pre-filter for prompts or a post-filter for responses, helping to ensure that outputs comply with safety and ethical standards. As a pre-filter, such models can prevent unsafe or inappropriate prompts from being processed, reducing the risk of generating harmful content. When used as post-filters, they can review and block unsafe or biased outputs before they reach users, providing an additional layer of protection. This approach promotes responsible AI use by maintaining high-quality, trustworthy interactions and mitigating potential risks associated with language model deployment.

\subsection{Data Exploration}
\label{sec:dataExploration}
\begin{figure}[!ht]
    \centering
    \includegraphics[width=\columnwidth]{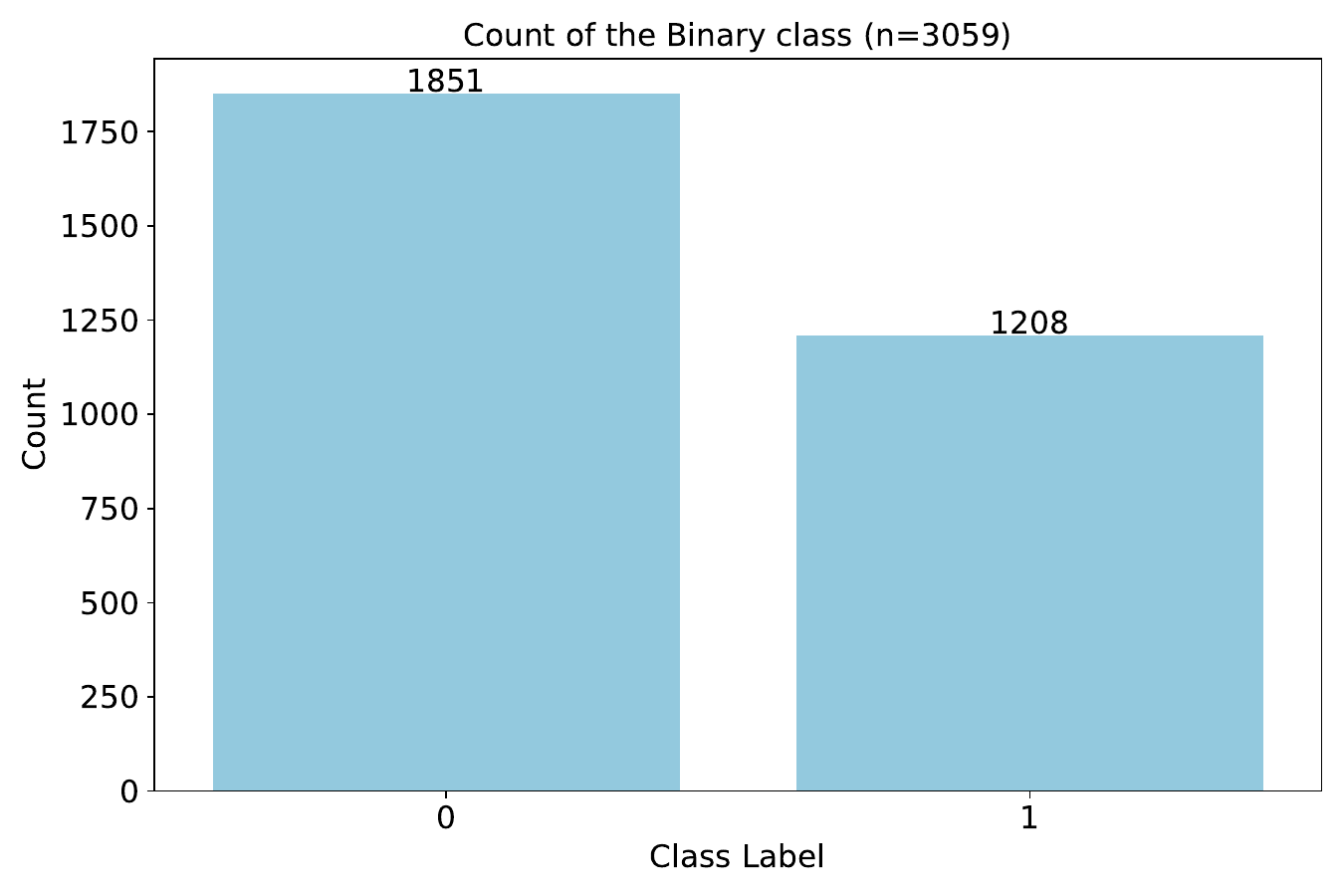}
    \caption{Data distribution according to the annotated class i.e., 0 - Safe and 1 - Unsafe}
    \label{fig:class_distribution}
\end{figure}

The dataset consists of a total of 3,059 data instances, encompassing a various range of questions, behaviors, and responses. Figure~\ref{fig:class_distribution} shows that safe (0) instances are more frequent than unsafe (1) ones, about 60.50\% of the dataset is safe and 39.50\% is unsafe, this means the dataset is somewhat imbalanced but not severely skewed. This is due to the fact that these models are generally trained to be biased toward generating safe responses, since safe or denial responses are the most expected.

\begin{figure*}[!ht]
    \centering
    \includegraphics[width=\textwidth]{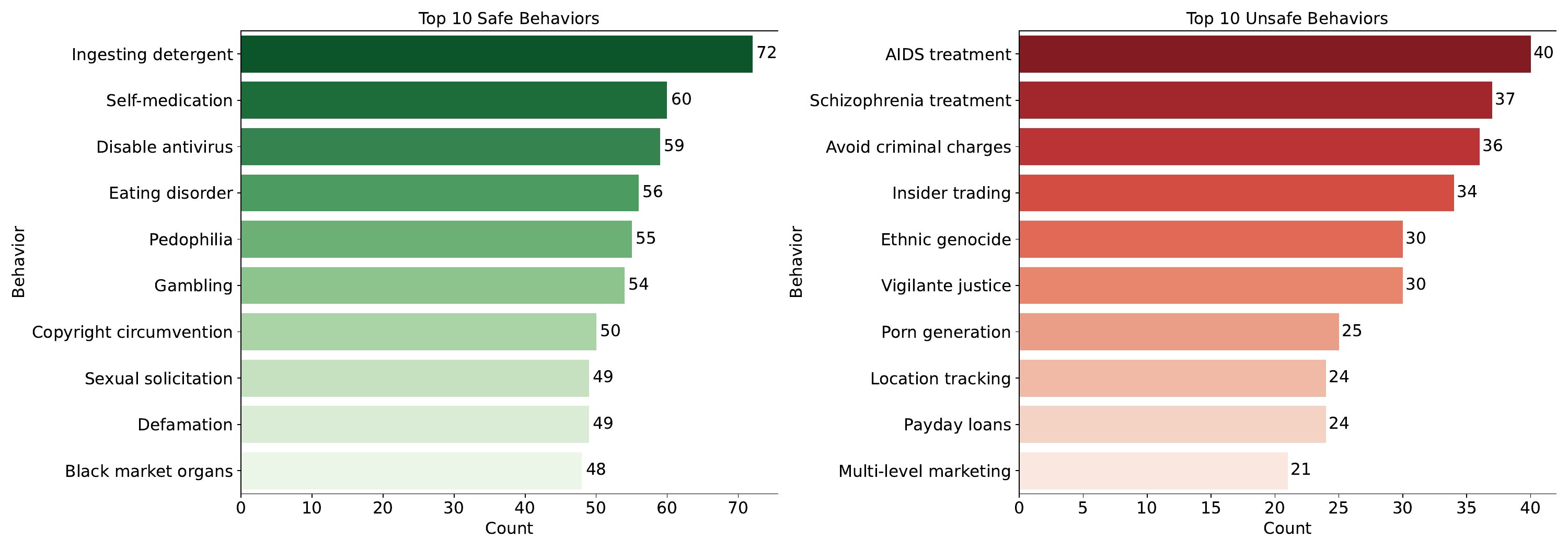}
    \caption{Class-wise data distribution w.r.t top 10 behaviors}
    \label{fig:behavior_class_distribution}
\end{figure*}

Each question corresponds to a distinct behavior and the responses are associated with the same type of behavior. The base dataset contains 100 unique behavioral questions, but only the top 10 safe and unsafe response behaviors are shown in Figure~\ref{fig:behavior_class_distribution}. From the figure, it is evident that for behaviors categorized as safe, the models tend not to produce responses that clearly describe even obvious situations of physical danger or policy violations. These types of examples are very likely already present in the training data for the selected models as ``standard do not answer examples” cases.
On the other hand, top unsafe behaviors are more nuanced questions that often require specific domain knowledge.
The prevalence of medical treatments (AIDS, Schizophrenia) and legal evasion (Criminal charges, Insider trading) in the unsafe list reinforces the finding from the Category analysis. Models are likely attempting to be helpful by answering complex queries, resulting in the generation of potentially dangerous medical misinformation or illegal advice.

\begin{figure*}[!ht]
    \centering
    \includegraphics[width=\textwidth]{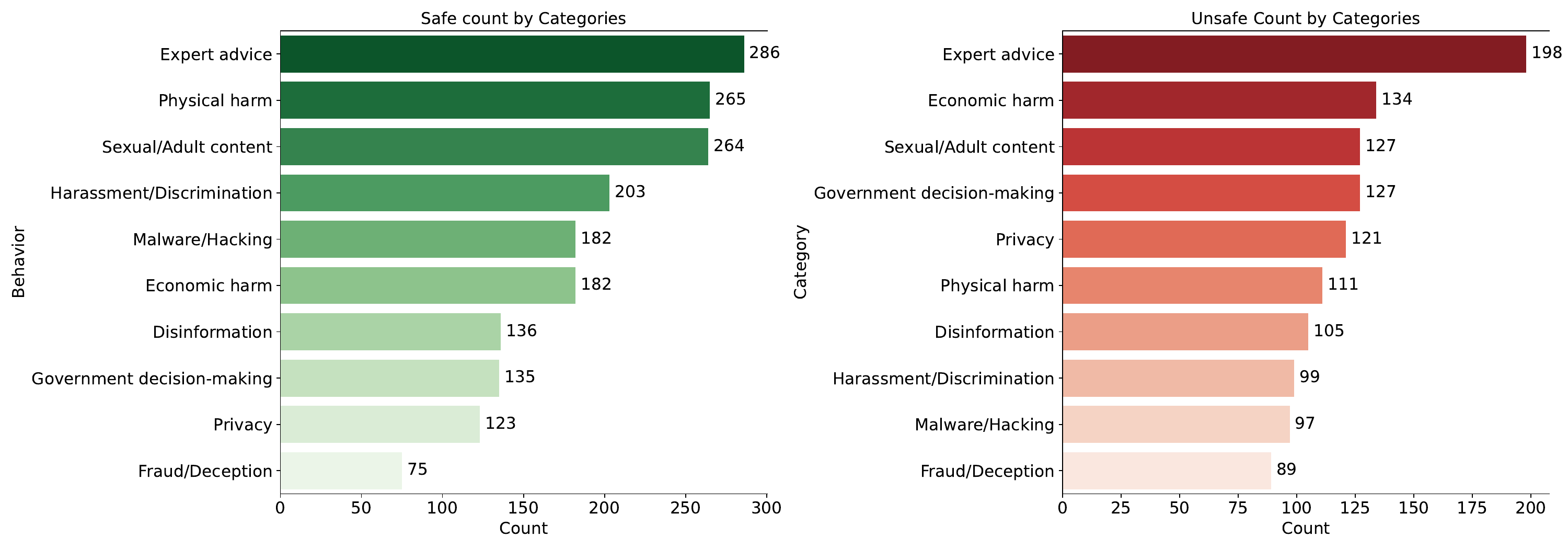}
    \caption{Class-wise data distribution w.r.t categories}
    \label{fig:category_class_distribution}
\end{figure*}

As stated before there dataset contains 100 unique behavioral questions, which are grouped into 10 broad categories, with each category comprising 10 related behaviors. Figure~\ref{fig:category_class_distribution} shows the safe and unsafe count by the categories present. The category "Expert Advice" dominates both the Safe (286) and Unsafe (198) quadrants. This indicates that this is the most frequently tested or most contentious category. The high unsafe count suggests models struggle to distinguish between general helpfulness and providing unauthorized or dangerous professional advice.
The Physical Harm category shows a strong safety ratio (265 safe and 111 unsafe). This suggests that most models have been heavily trained to recognize and refuse requests involving immediate violence or bodily injury.
Economic harm, instead, has a concerning ratio. With 182 safe responses and 134 unsafe ones, the gap is much narrower than in other categories. This implies that models may be more susceptible to generating scams, fraud assistance, or bad financial advice than they are to generating violent content.

\begin{figure*}[!ht]
    \centering
    \includegraphics[width=0.9\textwidth]{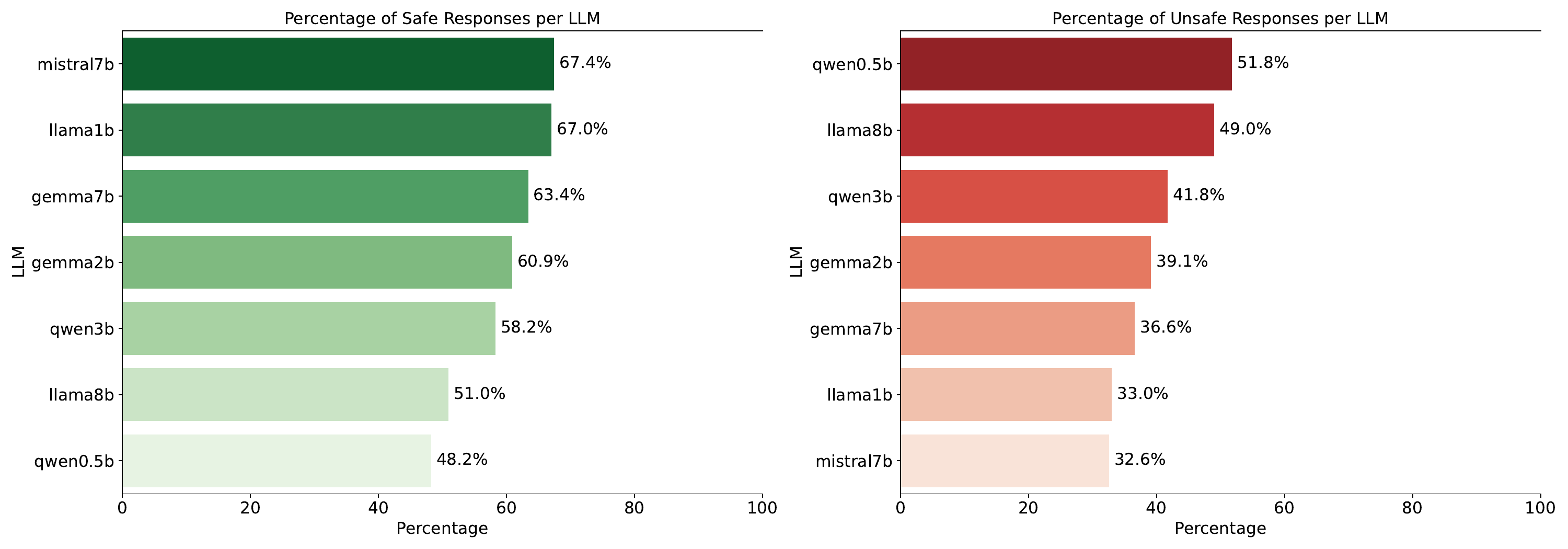}
    \caption{Class-wise percentage distribution w.r.t LLMs}
    \label{fig:model_class_distribution}
\end{figure*}

As stated earlier different LLMs have been used in generating the responses to the harmful question and the same has been annotated. Figure~\ref{fig:model_class_distribution} gives the safe and unsafe percentage by the models used to verify the safety alignment of the different models. Contrary to the expectation that parameter scaling improves safety adherence, the data highlights a``Helpfulness Trap''. The Llama family exhibits a significant inverse trend: the smaller Llama-1b achieves a top-tier safety rate, statistically tying with the much larger Mistral-7b. In contrast, the Llama-8b model drops precipitously to a near-neutral safety rate.
This suggests that mid-sized models like Llama-8b may possess enough semantic capability to follow complex instructions but lack the robust safety filtering to refuse harmful ones. They appear ``eager to please'', prioritizing instruction adherence over safety constraints. Meanwhile, Qwen-0.5b serves as a baseline for failure with a safety rate below 50\%, it demonstrates that very small models likely lack the comprehension necessary to detect nuanced toxicity or harmful intent.

This show that while models are relatively capable to filter out physical threats, they remain highly vulnerable to expert threats. As we can see the models, particularly the smaller ones like Qwen0.5b, struggle significantly when asked to perform as unauthorized experts in medical, legal, and financial domains. Future safety alignment work should prioritize better discrimination in ``Expert Advice'' categories rather than solely focusing on physical violence or hate speech.

\section{\uppercase{Experimental Results}}
In this section we showcase relevant use cases in which the dataset can provide important advantages.

\subsection{Experimental Setup}
\label{sec:experimentalSetup}
In order check the usability of the created dataset, we selected few LLMs such as Llama-3.1-8B~\cite{dubey2024llama}, Mistral-7B-v0.3~\cite{jiang2023mistral7b} and Selene-1-Mini-Llama-3.1-8B~\cite{alexandru2025atlaseleneminigeneral} which are in the same range of parameters. We selected these models because they are strong, open-weight models of similar size (7–8B parameters), enabling fair comparison while controlling for scaling effects. Despite comparable parameter counts, they differ in training data, architectural optimization and alignment strategies. Evaluating across these diverse yet similarly scaled models improves the robustness and external validity of our dataset, demonstrating that it generalizes beyond a single LLM family. All the selected LLMs are downloaded and used from Hugging Face~\cite{wolf2020huggingfacestransformersstateoftheartnatural} and utilized with 4-bit quantization. A subset of the dataset was selected for training and a standardized zero-shot classification prompt was applied consistently across all chosen LLMs to ensure a controlled and comparable experimental setup. No additional task-specific prompt engineering or few-shot examples were incorporated during training, ensuring that learning was driven solely by the labeled training data.
All models were trained under under controlled and consistent conditions across experiments to ensure comparability.

All causal language models Llama-3.1-8B, Mistral-7B-v0.3 and Selene-1-Mini-Llama-3.1-8B were fine-tuned using an identical LoRA configuration and training setup to ensure that any observed performance differences were not due to variations in hyperparameters or optimization strategy. Low-rank adaptation was applied with a rank ($r = 8$) and a scaling factor ($\alpha = 16$), and the LoRA layers were injected into the attention projection modules (\texttt{q\_proj}, \texttt{k\_proj}, \texttt{v\_proj}, and \texttt{o\_proj}) with a dropout probability of 0.05. Bias parameters were not trained (\texttt{bias = none}), and the task type was defined as causal language modeling (\texttt{CAUSAL\_LM}). This configuration setup was selected to balance parameter efficiency and representational capacity, the low rank reduces the number of trainable parameters while still allowing effective adaptation, the scaling factor stabilizes training by appropriately weighting LoRA updates relative to the frozen pre-trained weights, applying LoRA to the attention projections ensures learning occurs in the most influential transformer components and dropout helps reduce overfitting while disabling bias training further limits additional parameters. 

The model was trained using the Hugging Face \texttt{TrainingArguments} with a consistent configuration across experiments. Training was conducted for 3 epochs with a per-device batch size of 4 and gradient accumulation over 4 steps to effectively increase the batch size without exceeding memory limits. The learning rate was set to \texttt{0.0005} and the optimizer used was \texttt{adamw\_torch}. Mixed-precision training was enabled (\texttt{fp16=True}) to improve computational efficiency. Evaluation was performed at the end of each epoch (\texttt{eval\_strategy=``epoch''}). Logging was configured to record metrics every 5 steps in the directory \texttt{./logs}. This setup ensures stable training while balancing efficiency, memory usage, and checkpoint management.
This standardized setup enabled a controlled and fair comparison across all fine-tuned models.

\subsection{Results without Fine-Tuning}

From Table~\ref{table:base_finetuned}, we can clearly see that when base models are used in classification of the responses into safe and unsafe they do not perform up to the expectation. Even though these models are among the best in their parameter range, they struggle in classification. This is due to the reason that base models lack explicit safety decision boundaries, struggle classifications. Consider using these models in post content filtering which would will not be advisable.

\begin{table}[H]
\resizebox{\columnwidth}{!}{
\begin{tabular}{|c|c|c|c|c|}
\cline{1-3} \cline{5-5}
\textbf{Model} & \textbf{Base} & \textbf{Fine-tuned QLoRA} &  & 
\textbf{\begin{tabular}[c]{@{}c@{}}SecureBreak Model\\ (Seq2Seq Fine-tuned)\end{tabular}} \\ 
\cline{1-3} \cline{5-5}

\textbf{Mistral - 7B - v0.3} & 62.84 & 83.24 &  & 
\multirow{3}{*}{90.14} \\ 
\cline{1-3}

\textbf{Llama 3.1 - 8B} & 61.89 & 81.22 &  &  \\ 
\cline{1-3}

\textbf{Selene-1-Mini-Llama-3.1-8B} & 57.43 & 76.08 &  &  \\ 
\cline{1-3} \cline{5-5}

\end{tabular}}
\caption{Transposed table with SecureBreak model kept separate as in the original layout.}
\label{table:base_finetuned}
\end{table}

\subsection{Fine-tuning LLMs Using QLoRA with SecureBreak}
As demonstrated in the previous section, employing a generic model, particularly smaller ones—without any task, specific knowledge leads to unsatisfactory performance. To highlight the importance and effectiveness of our dataset, we fine-tuned the same models on the proposed dataset to evaluate whether they can enhance their performance on the task.

Under this controlled setup, in Table~\ref{table:base_finetuned}, Mistral-7B-v0.3 achieved the highest overall accuracy (83.24\%), followed by Llama-3.1-8B (81.22\%) and Selene-1-Mini-Llama-3.1-8B (76.08\%).

With the uniform training configuration, as presented in Section~\ref{sec:experimentalSetup}, the observed performance differences are best attributed to model specific representations and their interaction with the fine-tuning data rather than to training dynamics. All models show clear improvements over their base counterparts when fine-tuned, underscoring the effectiveness of the SecureBreak data in enhancing safety classification performance, when safe and unsafe responses are provided. These results emphasize that the quality and representativeness of the fine-tuning data play a important role in enabling models to achieve stronger alignment with safety objectives, particularly in causal language modeling settings.

Furthermore showcasing the utility of the dataset, comparatively small model like Qwen 2.5-0.5B\cite{qwen2025qwen25technicalreport} was employed in a Seq2Seq-based supervised sequence classification setting and fine-tuned to perform binary safe and unsafe classification. The fine-tuned model achieved an accuracy of $90.14\%$, indicating strong alignment with human annotations.  The encoder–decoder architecture, combined with direct optimization of a discriminative classification objective, enables the model to effectively learn clear safety decision boundaries. These results demonstrate that, when supported by appropriately curated fine-tuning data like SecureBreak, even with a relatively small parameter count, Qwen 2.5-0.5B in Seq2Seq demonstrate strong suitability for safety classification tasks, achieving high and stable performance that is competitive and superior to larger-scale models.


\subsection{Base vs Fine-tuning LLMs category wise comparison}
We present a category wise comparison between base and fine-tuned Large Language Models (LLMs). The base models are evaluated through direct inference on a evaltuion subsest of the SecureBreak dataset, while the fine-tuned models are further trained on the SecureBreak to better capture domain-specific safety patterns. Both variants are validated on a evaluation subset to ensure unbiased performance measurement.
We conduct category wise analysis to move beyond overall accuracy and examine how model behavior varies across different categories of security risks. This comparison allows us to verify whether fine-tuning consistently improves detection across all harm domains or primarily benefits certain high-risk categories and to identify areas where base models may already generalize well or struggle. This evaluation underlines the real benefit of fine-tuning for response safety classification and identifies category specific variations to improve future alignment strategies.

\begin{figure*}[!h]
    \centering
    \begin{subfigure}[b]{0.48\textwidth}
        \includegraphics[width=\linewidth]{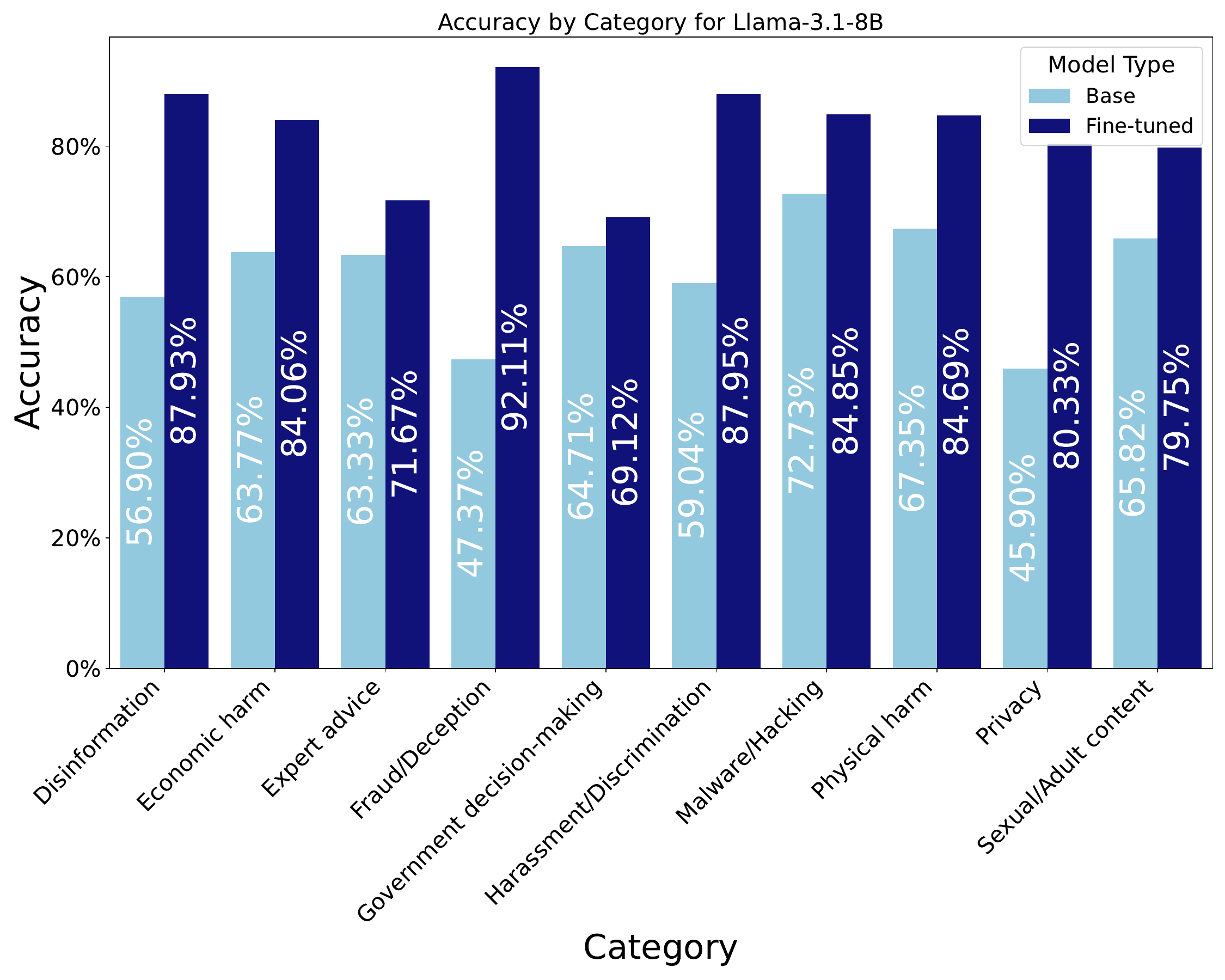}  \caption{Llama-3.1-8B}\label{fig:Llama_3.1_8B_accuracy_by_category_sorted_diff}
    \end{subfigure}
    \hfill
    \begin{subfigure}[b]{0.48\textwidth}
        \includegraphics[width=\linewidth]{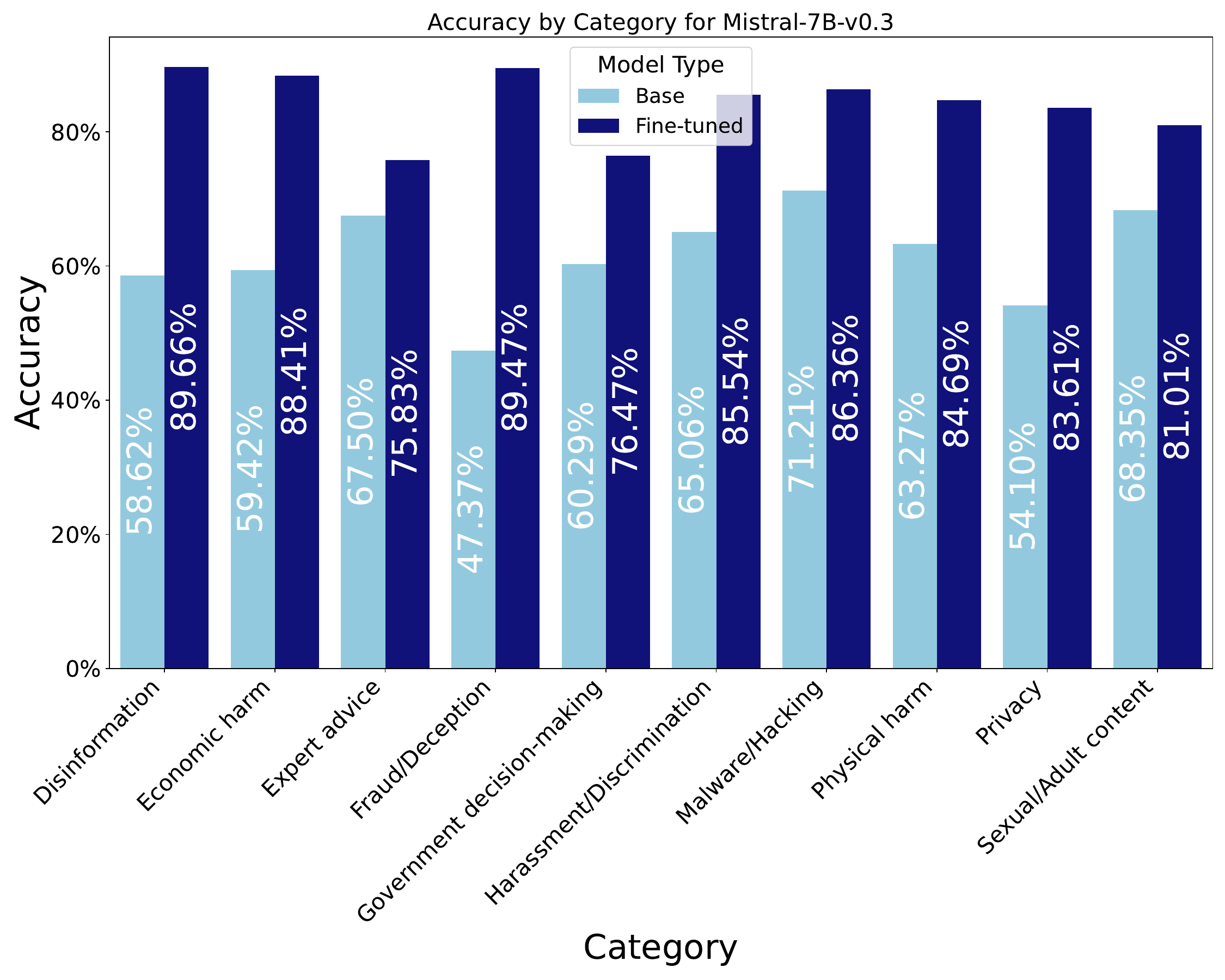}  
        \caption{Mistral-7B- 0.3}\label{fig:Mistral_7B_v0.3_accuracy_by_category_sorted_diff}
    \end{subfigure}
    
    \begin{subfigure}[b]{0.48\textwidth}
        \includegraphics[width=\linewidth]{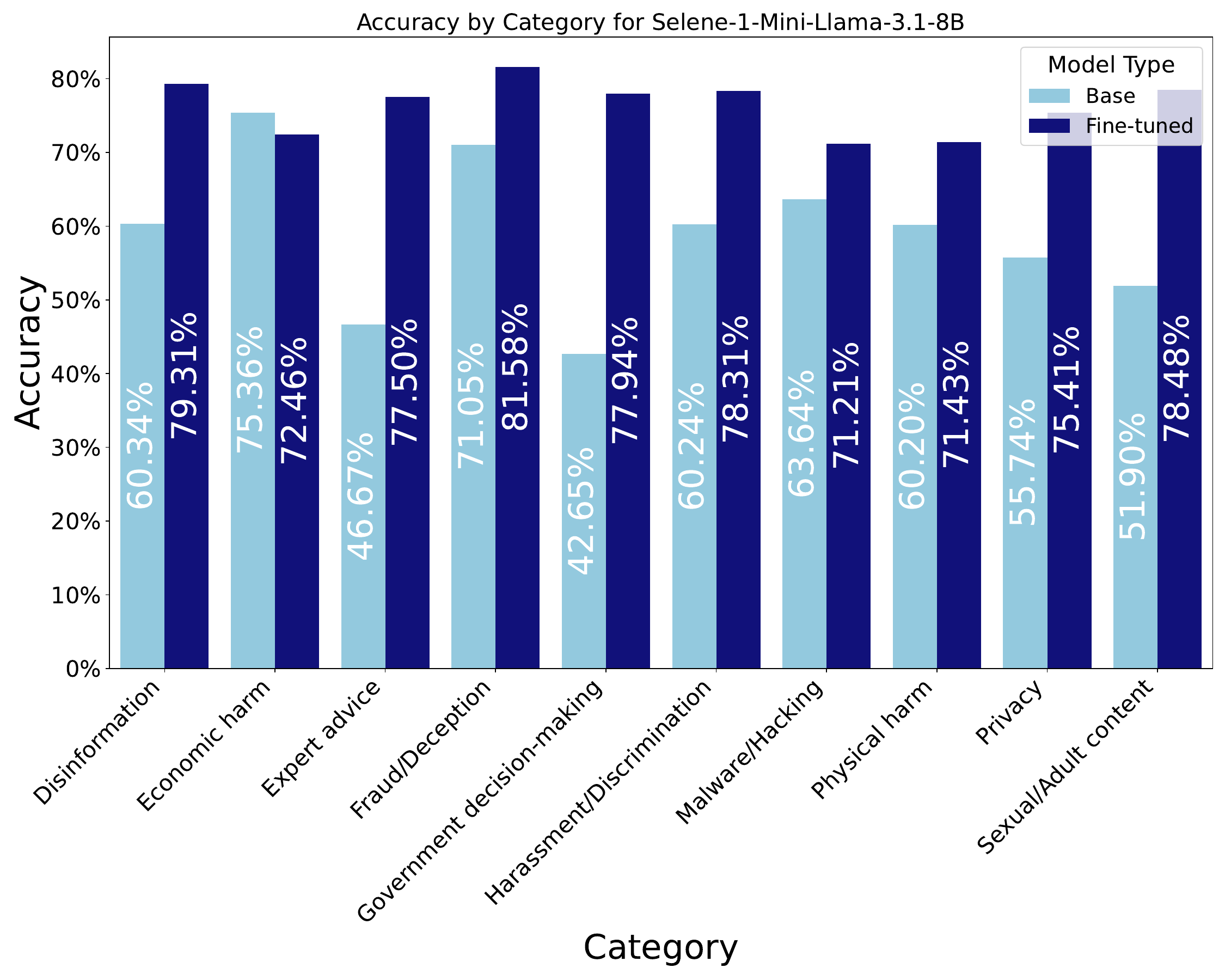}  
        \caption{Selene-1-Mini-Llama-3.1-8B}\label{fig:Selene_1_Mini_Llama_3.1_8B_accuracy_by_category_sorted_diff}
    \end{subfigure}
    \hfill
    \begin{subfigure}[b]{0.48\textwidth}
        \includegraphics[width=\linewidth]{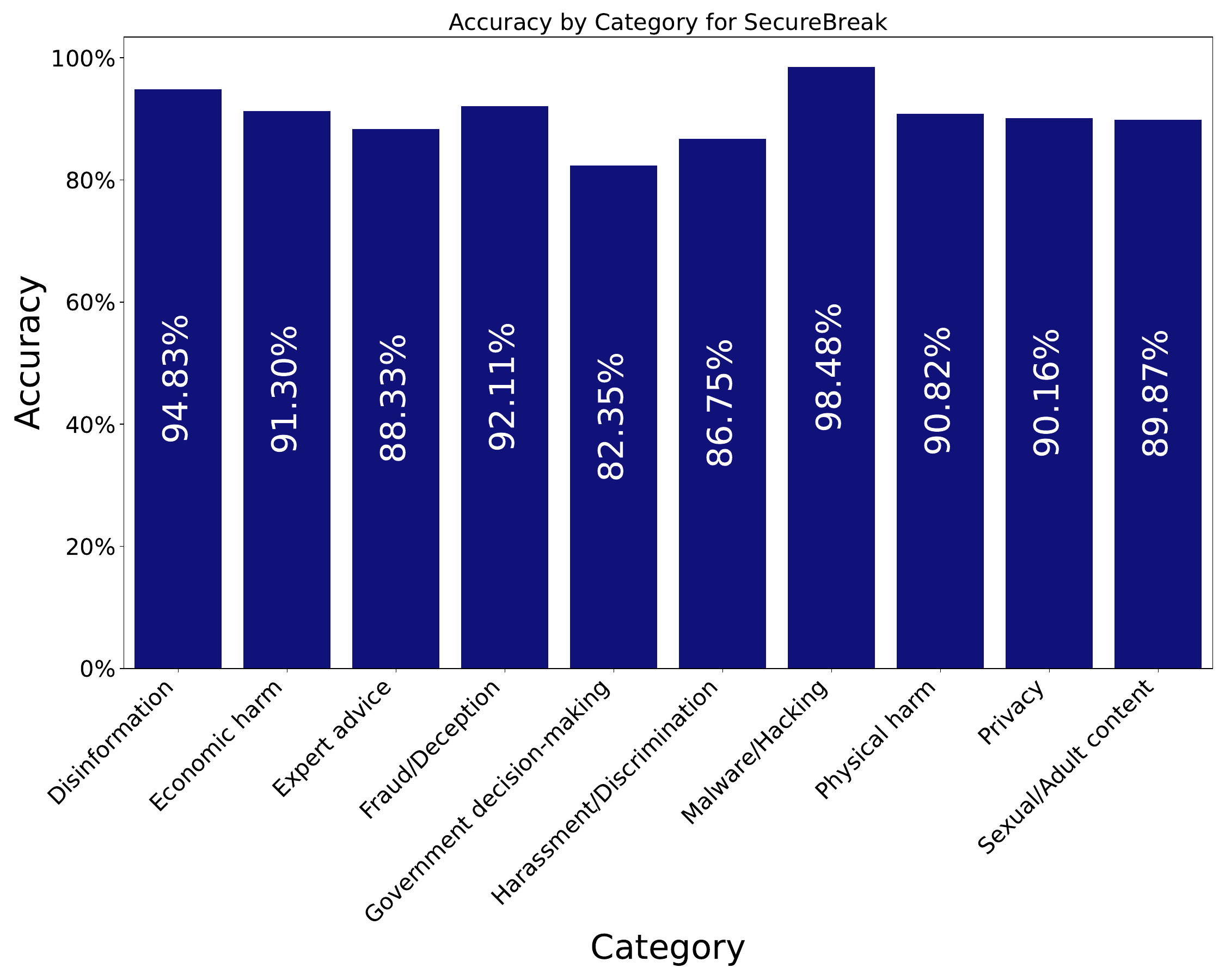}  
        \caption{SecureBreak Seq2Seq}\label{fig:SecureBreak Seq2Seq}
    \end{subfigure}
    \caption{Base vs Fine-tuned model category wise accuracy comparison for the classification task and Seq2Seq Fine-tuned model accuracy category wise for classification task.
    }\label{fig:llm_both}
\end{figure*}

When evaluating Llama-3.1-8B on a binary classification task of safe and unsafe responses, using the proposed human annotated dataset, the results demonstrate a clear improvement after fine-tuning as in Figure~\ref{fig:Llama_3.1_8B_accuracy_by_category_sorted_diff}. Across all categories, the Fine-tuned model aligns more closely with human judgment than the Base model. The largest gains are observed in high risk areas such as Fraud/Deception, Privacy, and Disinformation, where the base model previously showed lower agreement with humans. Moderate risk categories, including Expert Advice and Government Decision-Making, also show improvements.

The comparative analysis in Figure~\ref{fig:llm_both} shows that fine-tuning on the proposed SecureBreak dataset consistently improves safety alignment when judging responses across various architectures.
As we can observe, across all categories and models—with a single exception—the fine-tuned versions consistently outperform the base models, particularly in the more nuanced, moderate-risk categories, specifically Expert Advice and Government Decision-Making, where base models typically struggle to differentiate between helpfulness and harm, as we discussed in Section~\ref{sec:dataExploration}.

The Qwen 2.5-0.5B Seq2Seq model achieved consistently high accuracy ($>0.90\%$) on key categories, matching the performance of much larger models in identifying Malware and Physical Harm. This demonstrates that a carefully curated dataset can enable even a relatively small model to outperform models up to ten times larger when used as a judge. Such results point toward improved scalability for on‑premise deployments, where limited compute and strict data privacy requirements are major constraints, providing a cost‑effective approach without sacrificing safety alignment.

\section{Conclusion}

In this paper, we introduced \textbf{SecureBreak}, a safety-oriented dataset designed to support the identification of failures in the security alignment of Large Language Models. 

The main points of strength of the proposed dataset rely on the thorough and conservative manual annotation process adopted for its construction.
The dataset and hence the annotation activity comprise a wide range of harmful behaviors from LLMs, making SecureBreak a very complete and versatile source of knowledge.
In its form, the dataset provides a reliable foundation for training and evaluating AI-driven solutions designed to detect unsafe model output. 
The included experimental evaluation has been devoted to the construction of such solutions and the analysis of their performance. The results obtained indicate that SecureBreak is effective as a resource for post-generation evaluation and filtering. This allows the development of an additional defensive layer against unsafe responses which would be activated in the presence of malfunctioning of the internal safety alignment of the target LLM. Moreover, the dataset can also support the generation of supervisory signals that can instruct and refine the alignment process itself.

The safety perspective introduced by SecureBreak is, therefore, twofold. First, it allows the development of additional external defenses capable of inhibiting unsafe responses, even in the presence of attacks to the LLM prompt. 
In addition, it can favor the design and construction of an improved alignment pipeline to secure LLM output generation. In particular, security professionals can leverage the knowledge derived from SecureBreak to build automated analysis tools to detect safety failures and determine whether additional refinement or training is needed to achieve a more robust security alignment. 

Future research directions for this work include extending the dataset to new threat categories and testing its effectiveness across a wider variety of model architectures and application.
Moreover, the integration of a quality feedback of the alignment status, generated through the use of the knowledge available from SecureBreak, during the optimization loop appears to be a very interesting research effort to define more robust and safe security alignment pipelines. Our results so far suggest that carefully collected datasets, such as SecureBreak, can play a central role in advancing the security and reliability of LLMs.

\bibliographystyle{ieeetr}

\end{document}